\documentclass[runningheads]{llncs}
\usepackage{amsmath}  % load amsmath right after

\usepackage{enumitem} % For bullet points
\usepackage[T1]{fontenc}
\usepackage{makecell}
\usepackage{booktabs}  % For professional-looking tables

\usepackage{caption}          % load caption first
\usepackage{subcaption}       % now safe to use

\usepackage{float} % Allows use of the [H] specifier

\usepackage{cite}

\usepackage{graphicx}
\usepackage{xcolor}
\begin{document}
\title{`The World of AI': A Novel Approach to AI Literacy for First-year Engineering Students
}
%
% %\titlerunning{Abbreviated paper title}
% % If the paper title is too long for the running head, you can set
% % an abbreviated paper title here

\titlerunning{The World of AI} % Optional
% If the paper title is too long for the running head, set an abbreviated title here

\author{Siddharth Siddharth\inst{1}\orcidID{0000-0002-1001-8218} \and
Brainerd Prince\inst{1}\orcidID{0000-0003-4626-6536} \and
Amol Harsh\inst{1}\orcidID{0009-0007-9529-6850} \and
Shreyas Ramachandran\inst{1}\orcidID{0009-0009-3780-2174}}

\authorrunning{S. Siddharth et al.}
% First names abbreviated in running head. Use et al. if more than two authors.

\institute{Plaksha University, Mohali, India\\
\email{\{siddharth.s, brainerd.prince, amol.harsh, shreyas.ramachandran\}@plaksha.edu.in}
}
\maketitle              % typeset the header of the contribution
\begin{abstract}
This work presents a novel course titled `The World of AI' designed for first-year undergraduate engineering students with little to no prior exposure to AI. The central problem addressed by this course is that engineering students often lack foundational knowledge of AI and its broader societal implications at the outset of their academic journeys. We believe the way to address this gap is to design and deliver an interdisciplinary course that can a) be accessed by first-year undergraduate engineering students across any domain, b) enable them to understand the basic workings of AI systems sans mathematics, and c) make them appreciate AI’s far-reaching implications on our lives.
The course was divided into three modules co-delivered by faculty from both engineering and humanities. The planetary module explored AI's dual role as both a catalyst for sustainability and a contributor to environmental challenges. The societal impact module focused on AI biases and concerns around privacy and fairness. Lastly, the workplace module highlighted AI-driven job displacement, emphasizing the importance of adaptation.
The novelty of this course lies in its interdisciplinary curriculum design and pedagogical approach, which combines technical instruction with societal discourse. Results revealed that students’ comprehension of AI challenges improved across diverse metrics like (a) increased awareness of AI’s environmental impact, and (b) efficient corrective solutions for AI fairness. Furthermore, it also indicated the evolution in students' perception of AI's transformative impact on our lives.

% By integrating ethical perspectives with technical understanding of AI, the course aims to develop not only proficient engineers but also responsible global citizens.

% The use of collaborative teaching, blending STEM and humanities faculties, introduces students to critical, reflective thinking alongside technical problem-solving. This pedagogical innovation enables students to recognize the societal implications of AI and prepares them to the challenges of the 21st century.


\end{abstract}
\section{Introduction}
Artificial Intelligence (AI) is reshaping technology, society, and the workplace \cite{tai2020impact}, yet many first-year engineering students lack foundational knowledge of AI’s societal impact and ethical challenges, highlighting the need for comprehensive ethics training \cite{orchard2024analysis,feffer2023ai, goldenkoff2024left}. In response, the interdisciplinary ‘World of AI’ course was developed for first-year students; unlike traditional AI courses that emphasize technical and mathematical rigor, this course demystifies AI and focuses on its social implications, as research indicates that combining technical instruction with humanities perspectives enhances both AI proficiency \cite{educsci11070318} and ethical understanding \cite{usher2024aiethics}. Thus, the ‘World of AI’ course uniquely integrates STEM and humanities to equip students with essential technical skills and a critical awareness of AI’s broader societal impacts. The course comprises of three modules—AI and the Planet, AI and the Society, and AI and the Workplace—and introduces foundational AI concepts and ethics, emphasizing its dual nature as both a driver of innovation and a source of challenges, such as environmental stress \cite{patterson2021carbon}, societal biases \cite{pillsbury2022ai}, and workplace disruptions, thereby prompting a critical evaluation of its opportunities and limitations.  Collaboratively taught by STEM and humanities faculty, it addresses overlooked themes in traditional curricula and is evaluated via pre-course and post-course surveys and case
studies examining AI’s implications on the planet, society, and the workplace.

\section{Background}
Many first-year engineering students often lack foundational knowledge of AI and its broader societal implications when they begin their academic journeys. This gap in understanding has been documented across numerous studies, which highlight how traditional engineering curricula prioritize technical knowledge at the expense of societal and ethical perspectives \cite{baker2019educ}. Research has equally shown the negative impacts of developing and using AI technology.  Environmental implications
of AI have drawn increasing attention in recent years. Studies by Strubell \cite{strubell2019energy} emphasize the significant energy consumption associated with training large AI models, with data centers contributing substantially to global carbon emissions. These findings underscore the need for future engineers to recognize the environmental trade-offs of AI technologies and to innovate solutions that mitigate such impacts. However, students are neither taught nor exposed to these ecological costs of AI, thus leaving a crucial gap in their education. In the social domain, research by Obermeyer \cite{obermeyer2019dissecting} has illustrated how biases inherent in AI systems can perpetuate inequalities, particularly in sensitive areas like healthcare, criminal justice, and employment. This raises concerns about fairness and equity, which are often overlooked in traditional engineering curricula \cite{mehrabi2021survey}. 
% When AI models are used for tasks like facial recognition, predictive policing, or credit scoring, they can result in decisions that negatively affect communities of colour, even if the programmer did not intend to discriminate \cite{buolamwini2018gender,kirkpatrick2016battling,selbst2017disparate} . 
The work on the impact of AI on employment by Frey and Osborne \cite{frey2017future} argue that advancements in AI-driven automation have the potential to displace a significant portion of the workplace, particularly in industries reliant on routine tasks. At the same time, AI creates opportunities for new job roles, demanding a workplace that is adaptable and equipped with interdisciplinary skills. Despite these drastic recent shifts, engineering programs have not yet addressed how students can prepare for such disruptions, leaving them unprepared for the changing dynamics of the workplace \cite{autor2015why}. 

In response to this shortfall, several universities and online platforms now offer
stand-alone subjects on AI ethics or literacy, confirming rising demand yet
leaving first-year engineers underserved. Coursera’s open MOOC
\emph{Ethics of Artificial Intelligence}, delivered by Politecnico di Milano,
reaches a global audience but provides no project component and is not embedded
in any engineering curriculum \cite{courseraEthicsAI}.
Stanford’s flagship CS/ETHICSOC\,182 blends philosophy, public policy, and
computer science but requires CS\,106A, attracting mainly second- or
third-year CS students \cite{stanfordCS182}. By contrast, our \emph{World of AI}
is a required gateway in semester 1, co-taught by humanities and engineering
faculty, and uses case study and projects to couple foundational concepts with
environmental, social, and workplace analysis—an approach we have not found in
any other entry-level engineering programme.
% Existing literature highlights a consistent gap in equipping engineering students with the knowledge to address the societal, ethical, and environmental dimensions of AI. This lack of exposure leaves students underprepared to critically evaluate and engage with the transformative effects of AI on our lives.

\section{Methodology}

\subsubsection{Research Question}

The central research question addressed in this paper is: to what extent does this introductory course on AI impact the first-year engineering undergraduate students’ understanding about the role and impact of AI on the planet, the society, and the workplace?

\subsubsection{Course Design}

The course blended technical instruction with industry insights and interdisciplinary perspectives, guiding students from basic AI concepts to a critical evaluation of its societal implications through a series of progressive modules. Initially, three engineering lectures demystified key concepts like machine learning and neural networks, laying the groundwork for subsequent discussions, which were enriched by guest lectures from industry leaders and co-delivered sessions by engineering and humanities faculty. 

The curriculum was organized into three main modules: 
\begin{itemize}[]
    \item \textbf{AI and the Planet} This module examined AI’s dual role in driving sustainable progress while posing environmental challenges through high energy demands and extensive water use for data center cooling, and also highlighted AI’s potential to optimize renewable resource management.
    \item \textbf{AI and the Society} This module explored societal biases in decision-making, risks of biased AI model training, and the balance between personalization and privacy while emphasizing fairness, equity, and inclusivity.
    \item \textbf{AI and the Workplace} This module analyzed the evolving dynamics in an AI-driven economy, addressing both the emergence of new industries and the displacement of traditional roles, and underscored the need for reskilling initiatives to help workers adapt to technological change.
\end{itemize}

The course was delivered in person through two weekly 1-hour classroom sessions, supported by regular office hours with instructors and dedicated TA-led sessions for resolving doubts. This structure facilitated active engagement, consistent feedbacks.
% The detailed course outline is presented in Table~\ref{tab:course_outline}.

% \renewcommand{\arraystretch}{1.3} % Adjust row height for readability

% \begin{table}[h]
%     \caption{Course Outline} % Moved above the table
%     \label{tab:course_outline} % Label for referencing
%     \centering
%     \begin{tabular}{|p{4cm}|p{8cm}|} 
%         \hline
%         \textbf{Module} & \textbf{Topics} \\ 
%         \hline
%         AI and the Planet & 
%         \begin{enumerate}
%             \item Introducing AI and the Planet
%             \item Energy
%             \item Water
%             \item Human Cognition
%             \item The AI Challenge
%         \end{enumerate} \\
%         \hline
%         AI and the Society & 
%         \begin{enumerate}
%             \item Introducing AI and the Society
%             \item Painting Uniformity Over Diversity
%             \item Unraveling Bias and Prejudice in AI
%             \item Trust, Security, and Privacy Unveiled
%             \item Navigating the Ethics of AI
%             \item Charting a Responsible Path Forward
%         \end{enumerate} \\
%         \hline
%         AI and the Workplace & 
%         \begin{enumerate}
%             \item Understanding Employment
%             \item The Dual Impact of Technology on Employment
%             \item Changing Nature of Work
%             \item Human Role in an AI-driven World
%             \item Training for an AI-driven Future
%         \end{enumerate} \\
%         \hline
%     \end{tabular}
% \end{table}

\subsubsection{Data Collection}
This study employed a mixed-method, single-group pre-post design that combined quantitative multiple-choice surveys with qualitative open-ended responses. A total of 168 first-year engineering undergraduates (113 male, 55 female; ages 17–19) participated in a 14-week in-person elective course, with all participants providing informed consent for anonymized survey responses. Pre-course surveys, which included both multiple-choice and open-ended questions, established a baseline of students’ understanding of AI’s environmental, social, and workplace challenges and their ethical reasoning, probing their perceptions of AI’s contemporary impact. At the end of the course, a post-course survey was administered via Microsoft Forms during the final lecture, allowing students 15–20 minutes to reflect on their learning and to capture shifts in their knowledge and ethical perspectives after completing the modules. To support construct validity, all pre- and post-course questionnaire items were adapted from established instruments, including the Meta AI Literacy Scale (MAILS) \cite{mails2023} and UNESCO’s AI Ethics Readiness Assessment \cite{unesco2023readiness}. Ethical approval for the data collection was obtained from a research ethics committee.

In addition to the surveys, a rubric-based evaluation of the final group presentations measured learning outcomes; in these presentations, each group envisioned a future problem (by 2030) impacting the planet, society, or workplace and proposed an AI-based solution. The presentations were assessed on five criteria—Understanding of Future Problem (clarity in explaining the problem, its causes, and impact), Future Relevance (significance over the next five years), AI’s Contribution to the Problem (how AI might exacerbate or influence the issue), Corrective AI Solution (feasibility of using AI to address the problem), and Responsibility \& Safeguards (ethical risks and safety measures)---with each parameter rated on a 0-10 scale to enable performance comparisons across groups.
% For example, consider the analysis of the presentation submitted by Student Group 3 in our course.
% \begin{itemize}
%     \item \textbf{Understanding of Future Problem (10/10)}: Group 3 effectively chose and highlighted deepfakes as a major societal threat, emphasizing their role in misinformation and identity fraud.
%     \item \textbf{Future Relevance (9/10)}: They demonstrated how deepfake technology will become increasingly sophisticated and harder to detect.
%     \item \textbf{AI's Contribution to the Problem (10/10)}: Their analysis strongly showcased how AI-driven advancements in generative models amplify the issue.
%     \item \textbf{Corrective AI Solution (10/10)}: They proposed AI-driven detection frameworks and policy interventions to mitigate deepfake threats.
%     \item \textbf{Responsibility \& Safeguards (9/10)}: The group integrated ethical AI practices and regulatory oversight to combat deepfake misuse.
% \end{itemize}

\subsubsection{Preliminary Data Analysis}
The quantitative scoring methodology provided standardized measures across question types: for objective questions, correct responses earned 1 point and incorrect responses earned 0; in MCQs, each correctly selected option received 1 point while unselected or incorrect options scored 0; and scoring in Likert-scale responses was “Strongly Agree” = 5, “Agree” = 4, “Neutral” = 3, “Disagree” = 2, and “Strongly Disagree” = 1, with raw scores subsequently scaled to a 10-point scale. Qualitative responses were evaluated using an automated LLM-based method that compared each answer to an ideal response, normalizing similarity scores to a 10-point scale, with a sample of responses spot-checked by evaluators. Final presentations were assessed by an external evaluator using a five-category rubric, where each criterion was rated on a scale from 1 (poor) to 10 (excellent) and weighted equally.

\section{Results}

\begin{figure}[htbp]
    \centering
    \includegraphics[width=1\textwidth]{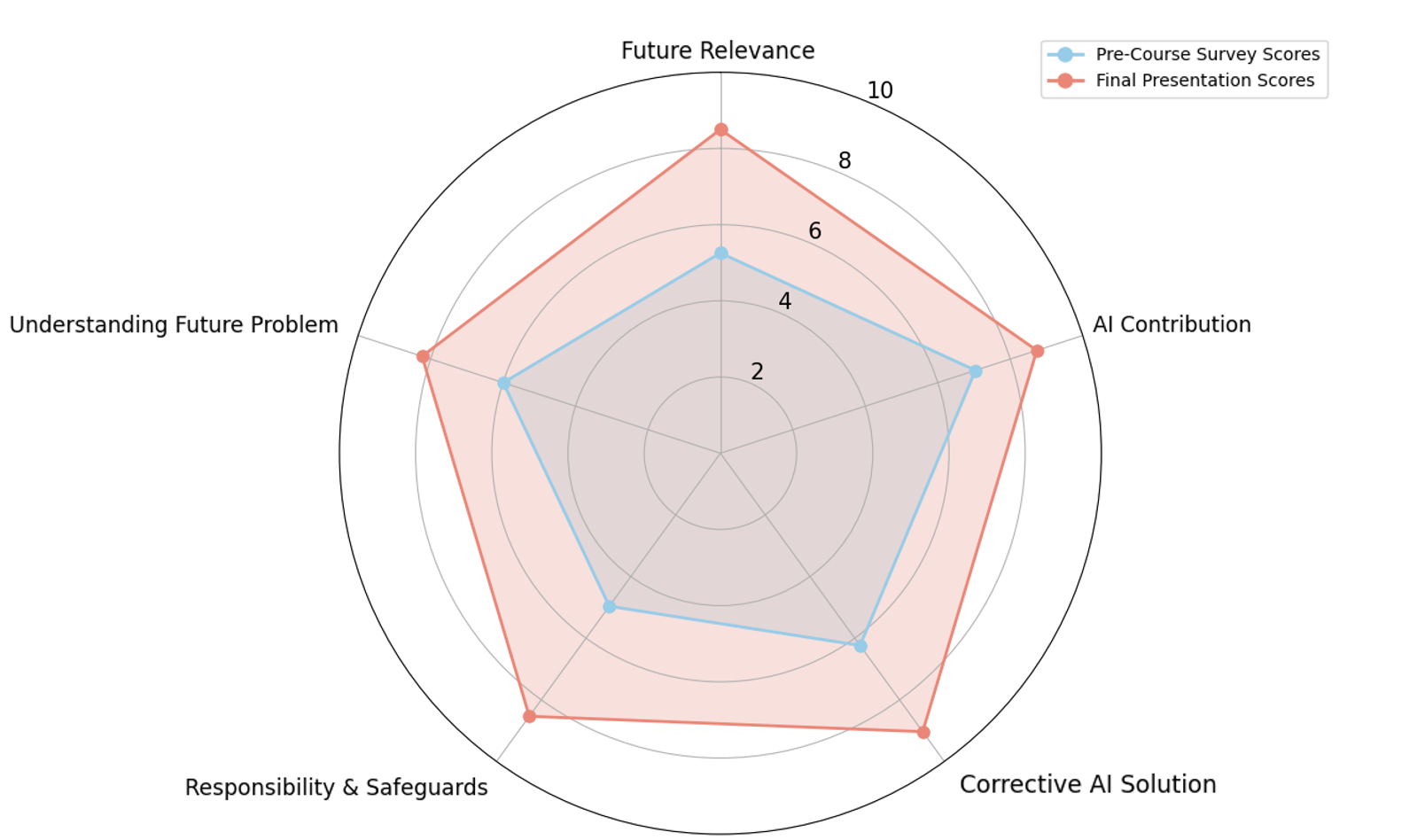}
    \caption{Radar Chart: Pre-Course Survey vs. Final Presentation Scores}
    \label{fig:radar_chart}
\end{figure}

Quantitative assessment of course effectiveness through pre-course survey vs. final presentation scores:
In order to systematically measure learning outcomes, we organized the pre-survey questions into five categories corresponding to the final project rubric: Understanding the Future Problem, Future Relevance, AI's Contribution to the Problem, Corrective AI Solution, and Responsibility \& Safeguards. These categories aligned with the rubrics outlined in the Methodology section, allowing for direct comparisons between students’ initial responses and their final presentation scores. Specifically, the “Understanding the Future Problem” category was linked to questions on reducing AI’s environmental impact and ensuring unbiased service. “Future Relevance” focused on balancing AI's benefits with sustainability by 2030, while “AI’s Contribution to the Problem” addressed energy requirements for data centers and the importance of prioritizing environmental impacts. “Corrective AI Solution” involved modifying an AI tool for monitoring pollution in diverse regions, and “Responsibility \& Safeguards” covered workplace bias and equitable AI-driven community recommendations. 

 In assessing various rubrics, students showed notable improvements from pre-survey to final presentation evaluations. By mapping these questions to the final rubric scores, we obtained a clear view of students’ progress across all five domains. For \textbf{Understanding the Future Problem (Rubric 1)}, an open-ended question addressing real-world AI challenges such as energy consumption and algorithmic bias, the mean score increased from 5.3 (\textit{SD} = 3.0) to 8.2 (\textit{SD} = 1.1). In \textbf{Future Relevance (Rubric 2)}, where students evaluated a future scenario emphasizing the urgency of sustainable AI within the next five years, scores improved from 7.0 (\textit{SD} = 2.3) to 8.5 (\textit{SD} = 1.5). \textbf{AI’s Contribution to the Problem (Rubric 3)} was gauged using two multiple-choice questions—one regarding the annual energy usage of AI-supporting data centers (correct answer: “Similar to global aviation industry energy consumption”) and another on the environmental impact of these centers—with normalized scores rising from 6.2 (\textit{SD} = 2.6) to 8.7 (\textit{SD} = 0.8). For \textbf{Corrective AI Solution (Rubric 4)}, which involved proposing enhancements to the AI tool’s coverage while ensuring fair data representation, the average score went from 5.0 (\textit{SD} = 3.1) to 9.0 (\textit{SD} = 0.9). Finally, \textbf{Responsibility and Safeguards (Rubric 5)}—assessed through two multiple-choice questions addressing potential biases in workplace areas (hiring, performance evaluation, employee monitoring, task automation) and equitable AI recommendations in urban planning (with the correct response being to expand data collection in underrepresented communities)—saw scores increase from 5.9 (\textit{SD} = 2.7) to 8.5 (\textit{SD} = 1.2), reflecting a strong overall gain in students’ awareness of fairness and ethical safeguards.

Because the gain scores were non-normal (Shapiro--Wilk, $p < .05$ for every rubric), we used paired Wilcoxon signed-rank tests ($W$ = test statistic, $n = 168$) and applied a Bonferroni correction to control the family-wise error ($p_{\text{adj}}$). All five learning outcomes improved significantly: \textit{AI Contribution} ($W = 1288.0$, $p_{\text{adj}} = 6.03 \times 10^{-13}$, $r = .47$), \textit{Responsibility \& Safeguards} ($W = 1012.0$, $p_{\text{adj}} = 1.21 \times 10^{-17}$, $r = .68$), \textit{Corrective AI Solution} ($W = 471.0$, $p_{\text{adj}} = 1.16 \times 10^{-22}$, $r = .77$), \textit{Future Relevance} ($W = 1717.0$, $p_{\text{adj}} = 5.30 \times 10^{-14}$, $r = .55$), and \textit{Understanding Future Problem} ($W = 954.0$, $p_{\text{adj}} = 7.57 \times 10^{-19}$, $r = .68$). Here, $p_{\text{adj}}$ is the Bonferroni-adjusted two-tailed probability, and $r$ is the rank-biserial effect size ($|r| \ge .50$ = large). Four rubrics show large practical gains ($r = .55$--$.77$) and one a moderate--large gain ($r = .47$), confirming that the observed improvements are both statistically robust and educationally meaningful.

\begin{figure}[htbp]
    \centering
    \includegraphics[width=0.8\textwidth]{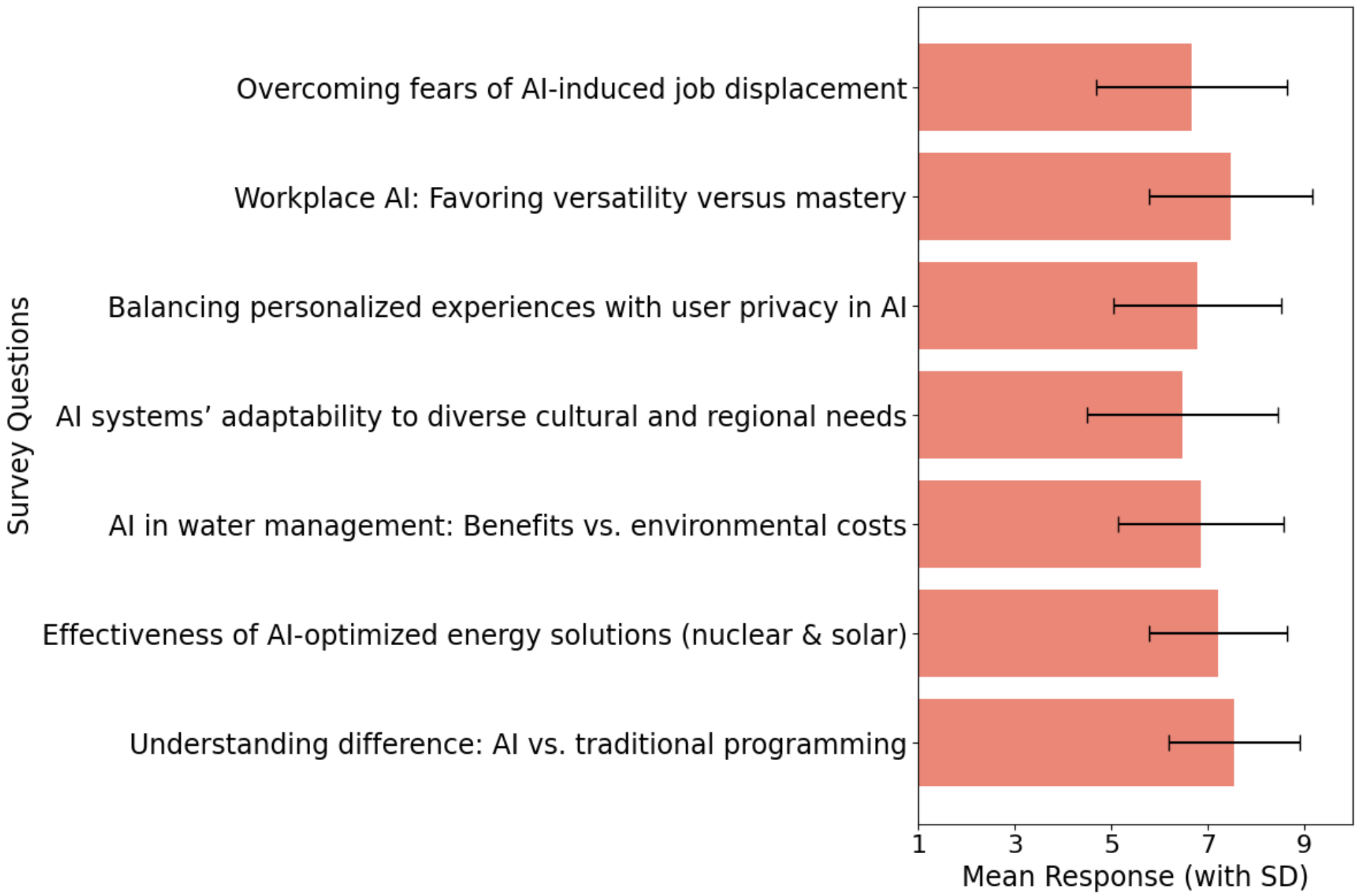}
    \caption{Students' post-course survey responses showing the extent of change in their understanding of various AI aspects (1 = Not at all, 10 = Completely).}
    \label{fig:post_survey_plot}
\end{figure}

Change in learners perspective from post-course survey:
The post-survey, illustrated in Figure \ref{fig:post_survey_plot}, captured detailed shifts in students’ understanding of AI across key areas—from technical distinctions between AI and traditional programming to its environmental, societal, and workplace impacts—through 7 Likert-scale questions that quantified changes since the course began. Questions addressed topics such as \textbf{AI \& the Planet}, asking to rate the effectiveness of AI-optimized energy solutions (e.g., nuclear for reliability and low-carbon output, solar for renewability despite intermittency) in balancing energy demands with sustainability; \textbf{AI \& the Society}, evaluating the ability of AI to balance personalized experiences with safeguarding user privacy; and \textbf{AI \& the Workplace}, assessing how effectively the course helped overcome fears of job displacement. The highest mean rating, approximately 7.5, was for understanding the difference between AI and traditional programming, while balancing personalization with AI systems’ adaptability received a mean score around 6.5, indicating moderate confidence. Overall, the responses reflect a positive shift in students’ perceptions across multiple AI domains.

\section{Discussion}

The central question of this study—how an introductory AI course impacts first-year engineering students’ understanding of AI’s role in the planet, society, and the workplace—is affirmatively answered through the course assessments in this research. Across all five rubrics, students demonstrated marked improvements in final presentations compared to baseline scores, while the post-course survey revealed moderate positive shifts in perceptions ranging from technical distinctions (machine learning versus traditional programming) to broader issues such as AI’s role in sustainability, with average ratings between approximately 6.5 and 7.5 on a 10-point scale (see Figure \ref{fig:post_survey_plot}). Nonetheless, several limitations warrant caution: the reliance on self-reported data introduces potential bias and subjective interpretation, and the absence of a control group or longitudinal follow-up leaves it unclear whether these attitudinal changes can be solely attributable to the course. Future work could explore longer-term impacts on students’ professional trajectories by tracking the cohort through subsequent semesters, internships, and early career experiences to better assess the enduring influence of the course’s conceptual foundations and ethical frameworks.

\begin{credits}
\subsubsection{\ackname}
The authors are thankful to Harish and Bina Shah School of AI \& CS at Plaksha University for providing the seed financial support for this research.

% \subsubsection{\discintname}
% It is now necessary to declare any competing interests or to specifically
% state that the authors have no competing interests. Please place the
% statement with a bold run-in heading in small font size beneath the
% (optional) acknowledgments\footnote{If EquinOCS, our proceedings submission
% system, is used, then the disclaimer can be provided directly in the system.},
% for example: The authors have no competing interests to declare that are
% relevant to the content of this article. Or: Author A has received research
% grants from Company W. Author B has received a speaker honorarium from
% Company X and owns stock in Company Y. Author C is a member of committee Z.
\end{credits}
\bibliographystyle{splncs04}
\bibliography{references} % This should match your .bib file name

\begin{thebibliography}{10}
\providecommand{\url}[1]{\texttt{#1}}
\providecommand{\urlprefix}{URL }
\providecommand{\doi}[1]{https://doi.org/#1}

\bibitem{autor2015why}
Autor, D.H.: Why are there still so many jobs? the history and future of workplace automation. Journal of Economic Perspectives  \textbf{29}(3),  3--30 (2015)

\bibitem{baker2019educ}
Baker, T., Smith, L.: Educ-ai-tion rebooted? exploring the future of artificial intelligence in schools and colleges. Tech. rep., NESTA (2019)

\bibitem{feffer2023ai}
Feffer, M., Martelaro, N., Heidari, H.: The ai incident database as an educational tool to raise awareness of ai harms: A classroom exploration of efficacy, limitations, \& future improvements. arXiv preprint arXiv:2310.06269  (2023), \url{https://arxiv.org/abs/2310.06269}

\bibitem{frey2017future}
Frey, C.B., Osborne, M.A.: The future of employment: How susceptible are jobs to computerisation? Technological Forecasting and Social Change  \textbf{114},  254--280 (2017)

\bibitem{goldenkoff2024left}
Goldenkoff, E., Cech, E.A.: Left on their own: Confronting absences of ai ethics training among engineering master’s students. In: Proceedings of the 2024 ASEE Annual Conference \& Exposition. American Society for Engineering Education (2024). \doi{10.18260/1-2--47724}

\bibitem{mehrabi2021survey}
Mehrabi, N., Morstatter, F., Saxena, N., Lerman, K., Galstyan, A.: A survey on bias and fairness in machine learning. ACM Computing Surveys  \textbf{54}(6),  1--35 (2021)

\bibitem{obermeyer2019dissecting}
Obermeyer, Z., Powers, B., Vogeli, C., Mullainathan, S.: Dissecting racial bias in an algorithm used to manage the health of populations. Science  \textbf{366}(6464),  447--453 (October 2019). \doi{10.1126/science.aax2342}

\bibitem{orchard2024analysis}
Orchard, A., Radke, D.: An analysis of engineering students' responses to an ai ethics scenario. Proceedings of the AAAI Conference on Artificial Intelligence  \textbf{37}(13),  15834--15842 (2024). \doi{10.1609/aaai.v37i13.26880}, \url{https://ojs.aaai.org/index.php/AAAI/article/view/26880}

\bibitem{patterson2021carbon}
Patterson, D., Gonzalez, J., Le, Q., Liang, C., Munguia, L.M., Rothchild, D., So, D., Texier, M., Dean, J.: Carbon emissions and large neural network training. arXiv preprint arXiv:2104.10350  (2021), \url{https://arxiv.org/abs/2104.10350}

\bibitem{pillsbury2022ai}
Pillsbury, W.S.P.L.: The impact of artificial intelligence on vulnerable populations in the workforce. Pillsbury Law  (2022), \url{https://www.pillsburylaw.com/en/news-and-insights/ai-impact-vulnerable-workforce.html}

\bibitem{courseraEthicsAI}
{Politecnico di Milano}: Ethics of artificial intelligence. \url{https://www.coursera.org/learn/ethics-of-artificial-intelligence} (2024), massive Open Online Course on Coursera, accessed 5 May 2025

\bibitem{educsci11070318}
Stadelmann, T., Keuzenkamp, J., Grabner, H., Würsch, C.: The ai-atlas: Didactics for teaching ai and machine learning on-site, online, and hybrid. Education Sciences  \textbf{11}(7) (2021). \doi{10.3390/educsci11070318}, \url{https://www.mdpi.com/2227-7102/11/7/318}

\bibitem{mails2023}
Straka, S., Latoschik, M.E., Wienrich, C.: {MAILS} — meta {AI} literacy scale: Development and testing of an {AI} literacy questionnaire based on well-founded competency models and psychological change- and meta-competencies. Computers in Human Behavior: Artificial Humans  \textbf{1}(2),  100014 (2023). \doi{10.1016/j.chbah.2023.100014}

\bibitem{strubell2019energy}
Strubell, E., Ganesh, A., McCallum, A.: Energy and policy considerations for deep learning in nlp. In: Proceedings of the 57th Annual Meeting of the Association for Computational Linguistics. pp. 3645--3650 (2019)

\bibitem{tai2020impact}
Tai, M.C.T.: The impact of artificial intelligence on human society and bioethics. Tzu Chi Medical Journal  \textbf{32}(4),  339--343 (2020). \doi{10.4103/tcmj.tcmj_71_20}

\bibitem{unesco2023readiness}
{UNESCO}: Readiness assessment methodology: A tool of the recommendation on the ethics of artificial intelligence. Tech. rep., United Nations Educational, Scientific and Cultural Organization, Paris (2023), \url{https://www.unesco.org/en/articles/readiness-assessment-methodology-tool-recommendation-ethics-artificial-intelligence}

\bibitem{stanfordCS182}
University, S.: Cs/ethicsoc 182: Ethics, public policy, and technological change. \url{https://web.stanford.edu/class/cs182/} (2025), course syllabus, accessed 5 May 2025

\bibitem{usher2024aiethics}
Usher, M., Barak, M.: Integrating ai ethics into science and engineering curricula: A case for explicit-reflective learning. International Journal of STEM Education  \textbf{11}(3),  45--60 (2024). \doi{10.1186/s40594-024-00493-4}

\end{thebibliography}
\end{document}